\documentclass[12pt]{iopart}

\usepackage{graphicx}
\usepackage{caption}
\usepackage{subcaption} 
\begin{document}
\title[Scatter from filter cavity optics]{Optical scatter of quantum noise filter cavity optics}

\author{Daniel Vander-Hyde$^1$, Claude Amra$^2$, Michel Lequime$^2$, Fabian Maga\~{n}a-Sandoval$^{1,3}$, Joshua R. Smith$^1$, Myriam Zerrad$^2$}

\address{$^1$Gravitational Wave Physics and Astronomy Center, California State University Fullerton, Fullerton, CA 92831, USA}

\address{$^2$Aix Marseille Universit\'{e}, CNRS, Centrale Marseille, Institut Fresnel, UMR 7249, 13013 Marseille, France}

\address{$^3$Department of Physics, Syracuse University, Syracuse, NY 13244, USA}

\ead{\mailto{danvande72hyde@csu.fullerton.edu}}

\begin{abstract}

Optical cavities to filter squeezed light for quantum noise reduction require optics with very low scattering losses. We report on measured light scattering from two super-polished fused silica optics before and after applying highly-reflective ion-beam sputtered dielectric coatings. We used an imaging scatterometer that illuminates the sample with a linearly polarized 1064\,nm wavelength laser at a fixed angle of incidence and records images of back scatter for azimuthal angles in the plane of the laser beam. We extract from these images the Bidirectional Reflectance Distribution Function (BRDF) of the optics with and without coating and estimate their integrated scatter. We find that application of these coatings led to a more than 50\% increase of the integrated wide-angle scatter, to $5.00\pm.30$\,ppm and $3.38\pm.20$\,ppm for the two coated samples. In addition, the BRDF function of the coated optics takes on a pattern of maxima versus azimuthal angle. We compare with a scattering model to show that this is qualitatively consistent with roughness scattering from the coating layer interfaces. These results are part of a broader study to understand and minimize optical loss in quantum noise filter cavities for interferometric gravitational-wave detectors. The scattering measured for these samples is acceptable for the 16\,m long filter cavities envisioned for the Laser Interferometer Gravitational-wave Observatory (LIGO), though reducing the loss further would improve LIGO's quantum-noise limited performance. 
\end{abstract}

\maketitle

\section{Introduction}

Optical filter cavities, formed by two or more highly reflective mirrors, provide frequency, amplitude, polarization, and mode geometry filtering and power buildup of resonating light fields. Such cavities have many useful applications, some being atomic clocks~\cite{2013Sci...341.1215H}, optical frequency combs~\cite{Steinmetz2009}, holographic noise meters~\cite{2012PhRvD..85f4007H}, and interferometric gravitational-wave detectors~\cite{Willke:98,Araya:97,Gossler:02,Fricke:12}. 

Without the implementation of squeezed light~\cite{Caves:85}, gravitational-wave detectors would eventually be quantum noise limited across a broad band, with radiation pressure dominating at lower frequencies and shot noise at higher frequencies~\cite{buonanno2000optical}. Squeezed light has recently been implemented, without filter cavities, in gravitational-wave detectors, improving their shot-noise-limited sensitivity~\cite{ligo:2011,aasi:2013}. Filter cavities provide a means of rotating the squeezing ellipse, as a function of frequency, to match the phase behavior of the detector and thereby to achieve a broadband reduction of quantum noise~\cite{PhysRevA.71.013806}. Research is underway to design and test filter cavities that will provide the correct phase behavior to accomplish this, without significantly degrading the squeezed states~\cite{Evans:13,Kwee:2014}. To achieve the optical filtering properties and noise performance required by advanced gravitational-wave detectors, such as the Advanced Laser Interferometer Gravitational-wave Observatory (Advanced LIGO)~\cite{Fritschel:2014}, these cavities must have long baselines, be isolated from seismic motion, and have very low optical loss, in particular, low scatter. 

Super-polished substrates with ion-beam-sputtered dilectric coatings of alternating layers of high and low index of refraction amorphous materials are the state-of-the art for low-loss optics~\cite{Harry:2011book,macleod2001thin,0264-9381-21-5-083}. Using such optics, filter cavities with very low loss, 5\,ppm per mirror, have been achieved~\cite{Isogai:13}. However, even these small levels of loss are enough to significantly degrade the quantum-noise reduction achievable using squeezed light especially when combined with other sources of technical noise and optical loss~\cite{Kwee:2014}.

It is not well known how the total scatter from such optics is distributed between surface microroughness, near-surface impurities and features, or scatter from coatings. 
To better estimate the scattering contributions of the coatings in quantum-noise filter cavities, we compared the light scattering properties of optics before and after the application of highly reflective ion beam sputtered coatings. The measured samples were two 2\,in diameter optics, very similar (same coating run, but substrates from different vendors) to those used for the filter cavity loss tests at MIT reported in Reference~\cite{Isogai:13}.

\section{Samples}

The two substrates investigated here were custom manufactured by Photon LaserOptik GmbH, with serial numbers E1200644-01 (Sample~1) and E1200644-02 (Sample~2). Both were 2\,in diameter 0.5\,in thick flat-concave fused silica, each with one -1\,m radius of curvature surface superpolished to rms $<$ 1\,\AA{} and one flat surface characterized by scratch dig 10-5. The flat surfaces of both optics, therefore, had larger surface roughness because they were not as highly polished. For the first set of measurements described below, both optics were uncoated. 

Prior to the second set of measurements, highly-reflective ion-beam sputtered coatings, formed by 40-60 layers of $\lambda/3$ and $\lambda/5$ stacks of SiO$_2$ and Ta$_2$O$_5$ (exact design proprietary), were deposited on top of the curved surfaces, along with anti-reflection coatings on the flat surfaces, by the coating manufacturer Advanced Thin Films. After coating, the transmissivity of Sample~1 and Sample~2 were measured to be 192 $\pm$ 1.3\,ppm and 1.55 $\pm$ .14\,ppm, respectively~\cite{Isogai:13}. 

Each sample is marked on its barrel with its serial number and an arrow pointing toward the curved surface. Care was taken to orient the samples such that the arrow pointed toward the laser beam, however, the rotation of the samples about their optical axis was not consistent between measurements.  

\section{Setup}

\begin{figure}[ht]
\centerline{\includegraphics[width=.8\columnwidth]{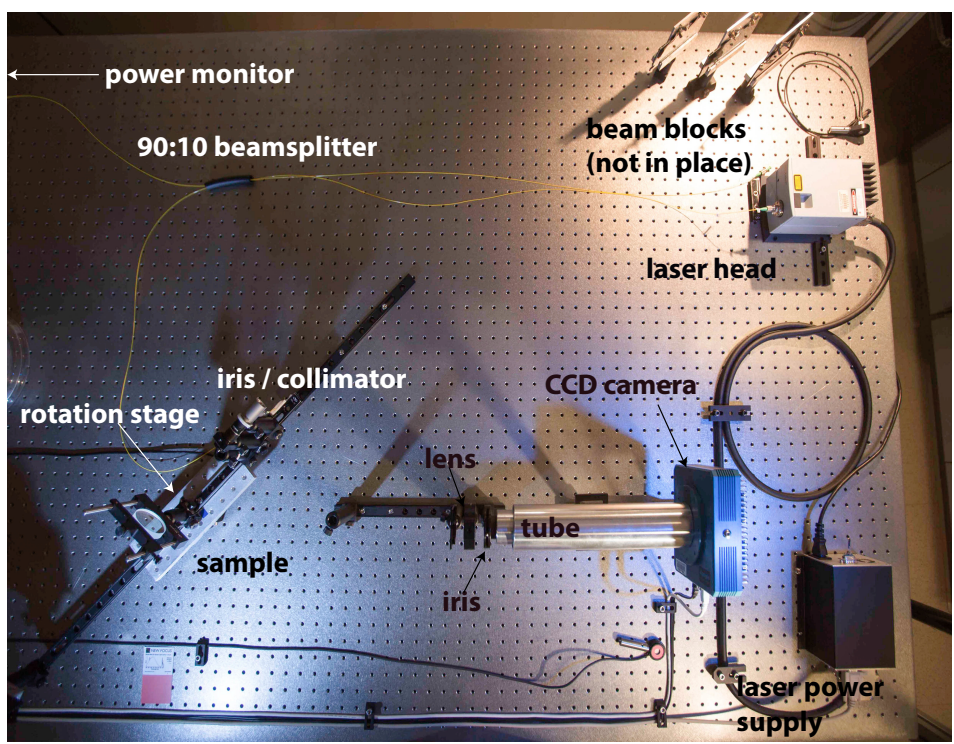}}
\caption{Overhead view of the imaging scatterometer. The sample is illuminated at a fixed angle of incidence by 1064\,nm laser light emerging from the iris and collimator. Both the sample and laser optics are mounted on a rotation stage. The CCD camera images scattered light from the sample at discrete azimuthal scattering angles from 0 to 90 degrees in the plane of the laser beam. During measurements, the beam blocks are attached to the rails mounted on the rotation stage to form traps for the specular reflected and transmitted beams.}\label{fig:scatterometersetup}
\end{figure}

The imaging scatterometer used to characterize the light scattering behavior of the two samples before and after coating is described in references~\cite{Magana-Sandoval:12, Padilla:14}, and shown in Figure~\ref{fig:scatterometersetup}. The basic operation of the scatterometer is that the sample is illuminated by laser light at a fixed incident angle, and the sample and laser launch are rotated together on a stage, allowing a fixed camera to image the scattering through a range of angles. 

This setup uses horizontal linearly polarized 1064\,nm wavelength light from a (Innolight Mephisto INN401) ND:YAG laser source. The laser is coupled into a fiber beam splitter with 90\% going toward the setup and 10\% used for power monitoring. The light emerging from the 90\% fiber further passes through a reflective collimator, which prepares an output beam diameter of roughly $8$\,mm, a horizontal polarizer to further improve its linear polarization, and an iris that truncates the beam that is incident on the sample to a width of $3.9\pm0.1$\,mm. The incident power ranged from roughly 50 to 80\,mW for the four runs (two samples, coated and uncoated), and had a power stability of $\pm$1\,mW over typical measurement times in a given run. 

For these measurements, the beam was somewhat non-uniform (as can be seen in the images in Figure 2) due to poor centering through the incident beam's iris. However, the non-uniform beam shape should not affect the values presented below, since those are based on total scattered power at a given angle versus total incident power, summing over the beam shape\footnote{A well controlled or well measured incident beam shape would allow a better visual or computer interpretation of the scattering images, through normalization of the images versus incident irradiance. To allow for this, we have since purchased a beam analyzer to be used in future scattering measurements.}.

A ($f=200$\,mm) converging lens and a second iris are used to image a roughly 1.5x1.5 inch region of the sample on to a 1024x1024 pixel (Apogee Alta U6) CCD-camera. An aluminum tube with a band-pass optical filter at its entrance is also used to reduce room light. To reduce stray light, the specularly reflected and transmitted beams are dumped in traps made from shade 10 welder's black glass mounted to the rails extending from the rotation stage. The fiber launch and sample, with a small fixed incident angle (4 degrees, to allow directing the reflected light to a beam dump), are mounted on a rotation stage. The stage is rotated in one degree increments from 0 to 90 degrees, stopping at each for the camera to take an image from that viewing angle. The automation is controlled by a LabView program and measurements are made in a clean room environment with no humans present. 

Once the images have been collected over the desired range of angles, a MatLab script is used to subtract dark images from them, sum the pixel values in the region of interest, and multiply by a calibration factor determined in a separate set of measurements on a scattering standard, following the procedure in reference~\cite{Magana-Sandoval:12}. The primary output of this is the standard  bidirectional reflectance distribution function~\cite{Stover:2012book} that quantifies the amount of scattered light. 

\begin{equation} 
\mathrm{BRDF}(\theta_s) = \frac{P_s}{P_i \: \Omega \: cos(\theta_s)},
\end{equation} 

where $P_i$ is the incident power on the sample, $P_s$ is the scattered  light power measured by the detector, which subtends by a solid angle $\Omega$ and is viewing from a discrete scattering angle, $\theta_s$, in the plane of the laser beam. 

For these images, even after subtracting dark images, a strong linear dependence of the BRDF on the area enclosed by the region of interest chosen suggested that there was a significant constant background of light in the images. To account for this, BRDF was calculated for a series of regions of interest in each image (the smallest of which are those shown in Figure~\ref{fig:rois}), each centered on the same area, but progressively larger (though not going past the sample's surface in the images). The data, BRDF versus area enclosed by the regions of interest, were then fit with a line, whose slope indicated the background light level and whose y-intercept represented the true underlying BRDF, i.e., the light scattered from the incident beam at the sample, which should be a constant, not varying with enclosed area for regions of interest that fully enclose the beam. Because of the collective influence of many non-zero pixels, the background is not small. Background would have added an additional 1--8 ppm to the integrated scatter values reported in Table~\ref{tab:tis}.

\section{Measurements}

\begin{figure}
\centering
\begin{subfigure}{.5\linewidth}
\includegraphics[width=\linewidth]{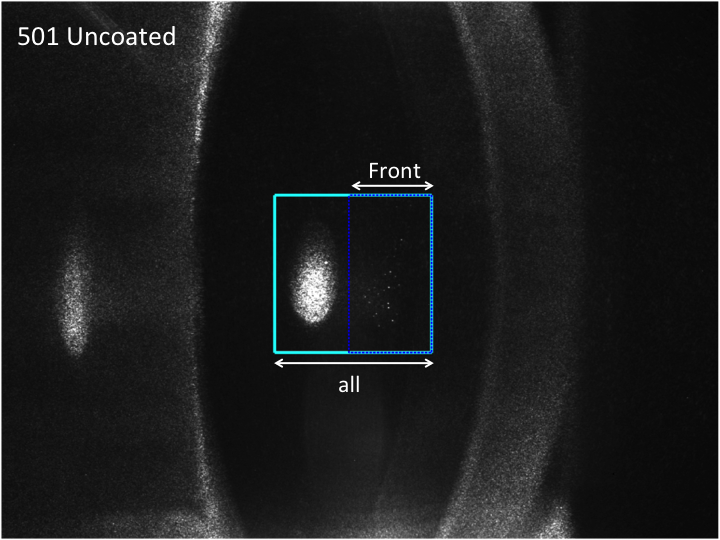}  
\label{fig:sub1} 
\end{subfigure}%
\begin{subfigure}{.5\linewidth}
\includegraphics[width=\linewidth]{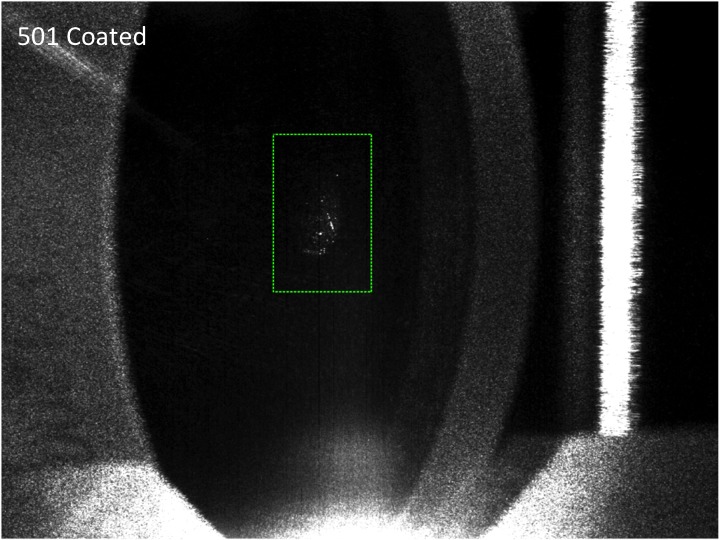}
\label{fig:sub2}
\end{subfigure}
\begin{subfigure}{.5\linewidth}
\includegraphics[width=\linewidth]{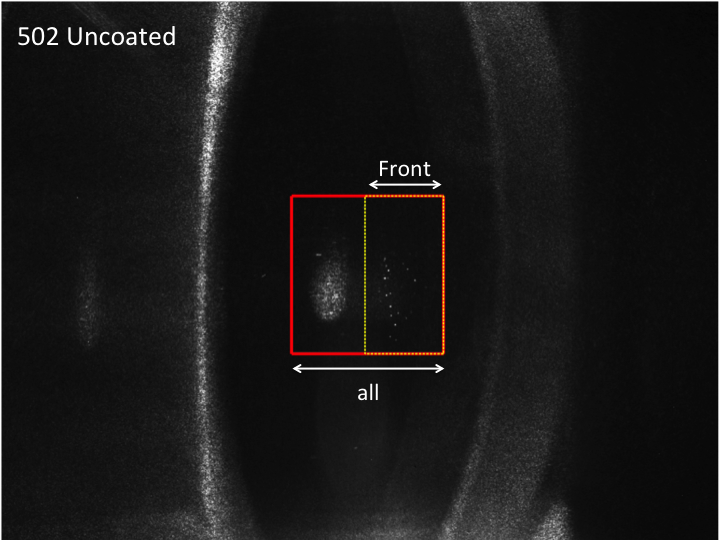}  
\label{fig:sub1} 
\end{subfigure}%
\begin{subfigure}{.5\linewidth}
\includegraphics[width=\linewidth]{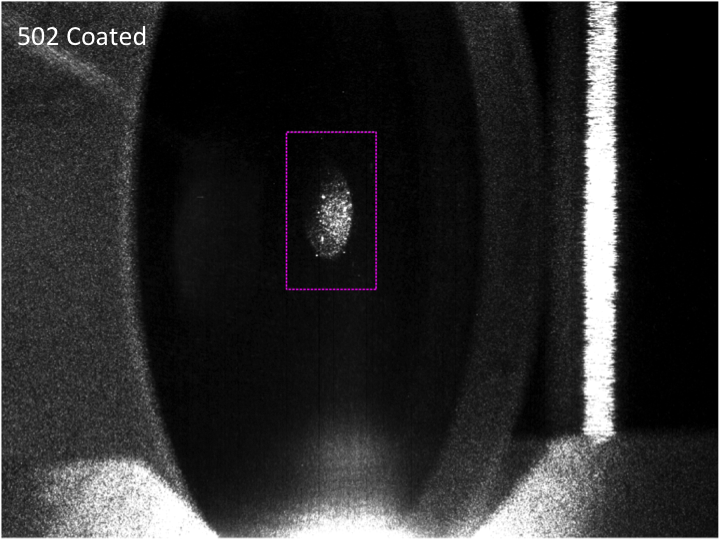}
\label{fig:sub2}
\end{subfigure}
\caption{Scattered light from Samples 1 (above) and 2 (below) at $\theta_s = 63^{o}$, before (left) and after (right) the high reflectivity coatings were applied. Rectangles show the regions of interest used to calculate BRDF.}\label{fig:rois}
\end{figure} 
 
Images of the illuminated samples were taken by the scatterometer for viewing angles from $0^{\circ} \le \theta_s \le 90^{\circ}$. However, in the analysis angles smaller than $11^{\circ}$ are excluded because the CCD camera's view of the optic is partially obstructed and some stray light from the specular reflection contaminates the images. Similarly, angles larger than $77^{\circ}$ are excluded because for these angles the view of the sample's surface is very compressed, making it difficult to distinguish scatter from the surface from that from the unpolished barrel of the optic and the sample holder. The exposure times of the images were chosen to be as long as possible (for better signal-to-noise) without CCD saturation in the regions of interest. Consequently exposure times varied from 15 to 65 seconds. Dark images, with the laser turned off, but the sample and rotation stage at the same values as for the scattering images, were taken and subtracted from the scattering images to reduce ambient light and CCD image defects.  

\begin{figure}[h]
\centerline{\includegraphics[width=1\textwidth]{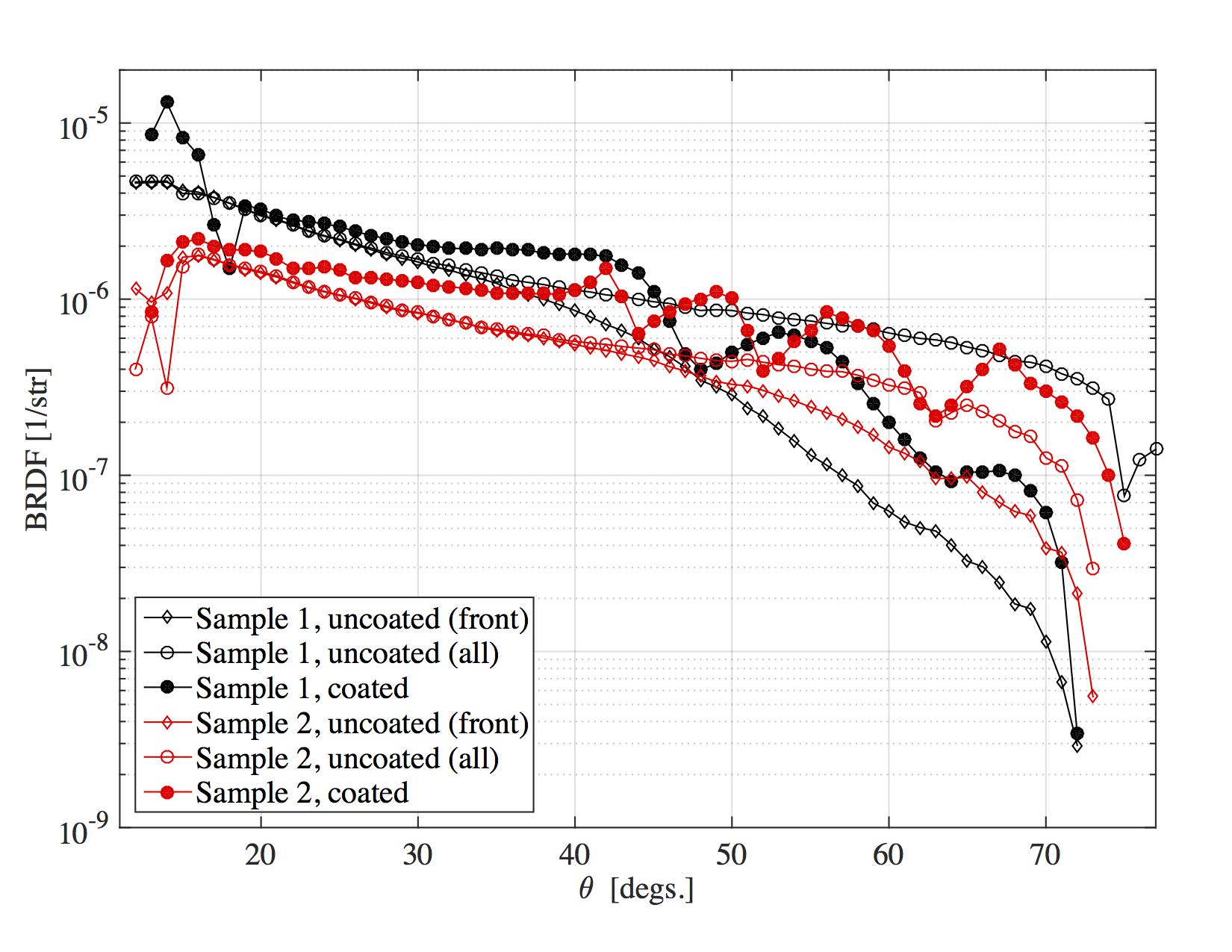}}
\caption{BRDF values versus scattering angle for Samples 1 and 2 corresponding to the regions of interest shown in Figure 3. After coating, the front/curved surfaces of the samples have higher BRDF at nearly all angles, and show a strong series of humps and zeroes.}\label{fig:BRDFplot}
\end{figure}

Figure~\ref{fig:rois} shows images of the scattered light from the coated and uncoated samples at $\theta_s = 63^{o}$.  For the uncoated samples, bright scattering is observed from the flat surfaces (left spot in the images) because they have standard, rather than super-polished surfaces. Much fainter scatter is also visible from the curved surfaces (right spot for uncoated samples), showing a combination of speckle/glow characteristic of surface microroughness, and bright points characteristic of point defects and contaminants\footnote{Research is underway, but not yet completed, to properly assess the budget of point scatter and glow/speckle from such images.}. The flat surfaces were not of interest in this study, but they did complicate the analysis by overwhelming scatter from the curved surfaces for small $\theta_s$. In order to better separate the contribution of the front, curved surfaces, two different regions of interest were defined for the uncoated samples. To attempt to separate the scattered light contribution of the front/curved surface from the total scatter (front surface, back surface, and bulk scattering), two separate regions of interest, 'front' and 'all', were defined as above for each image of the uncoated optics. Because these regions overlap at small angles, scatter from the front/curved surface could only be entirely isolated for large angles ($\theta_s > 50^{\circ}$, responsible for the separation between 'front' and 'all' curves seen in Figure~\ref{fig:BRDFplot}). 

Figure~\ref{fig:BRDFplot} shows the BRDF values measured for the two samples. The BRDF for the uncoated samples are relatively featureless and unfortunately dominated by the surface roughness of the flat surfaces for angles below about $45^{\circ}$. At higher angles, where the spot on the front/curved surfaces no longer spatially overlaps with scatter from the back/flat surface, the BRDF is quite low, in the $10^{-7}$--$10^{-8}$\,str$^{-1}$ range. Based on this, it seems likely that the BRDF of the front/curved surfaces should be several times lower than the curves shown at smaller angles. 

Both samples exhibited a significantly higher BRDF curve after coating, except in narrow dips, where the coated BRDF nearly reaches the uncoated BRDF. In addition, the BRDF curves for the coated sample show a strong pattern of maxima and minima with the minima at roughly $48^{\circ}$ and $63^{\circ}$ for Sample~1 and  $44^{\circ}$, $52^{\circ}$, and $63^{\circ}$ for Sample~2. This is seen in the images (and videos) as a modulated brightness of the speckle-glow pattern on the curved mirror surface and its origin is discussed further in the next section.



 
 \begin{table}[ht]
  \begin{center}
    \begin{tabular}{lcccc}
    \hline
    Optic & Angles (degrees) & $\Omega$ (sr) & IS (ppm) & Increase\\
    \hline
Sample 1, uncoated (all) & $11-76$ & 1.48$\pi$ & $4.94 \pm .30$ & \\
Sample 1, uncoated (front) & $11-76$ & 1.48$\pi$ & $3.27 \pm .20$ & \\
Sample 1, coated & $11-76$ & 1.48$\pi$ & $5.00 \pm 0.30$ & 52 \%\\
\hline
Sample 2, uncoated (all) & $12-77$ & 1.51$\pi$ & $2.21 \pm .13$ & \\
Sample 2, uncoated (front) & $12-77$ & 1.51$\pi$ & $1.91 \pm .11$ & \\
Sample 2, coated & $12-77$ & 1.51$\pi$ & $3.38 \pm .20$ & 77 \%\\
    \hline
    \end{tabular}
  \end{center}
\caption{Integrated scatter (IS) estimates for Sample~1 and 2, for the regions of interest in Figure~\ref{fig:rois}, before and after coating. These were calculated by integrating the cosine-corrected BRDF values over the azimuthal angles measured and assuming independence on polar angle. Also shown is the percentage increase in the IS after coating, compared to the pre-coating IS for the front/curved surface.}
\label{tab:tis}
\end{table}

  

Following Stover~\cite{Stover:2012book} and the procedure in reference~\cite{Padilla:14}, total scatter was estimated by integrating the BRDF values in Figure~\ref{fig:BRDFplot} times $\mathrm{cos(}\theta\mathrm{)}$, over the measured azimuthal angle ranges and all polar angles (assuming independence on polar angle). The results are shown in Table~\ref{tab:tis}. These values indicate that the application of highly reflective coatings on the curved surfaces lead to an increase in scatter by more than 52\% and 77\% (since the uncoated data is tainted by scatter from the rough back/flat surfaces).




\section{Model}
 
\begin{figure}[h]
\centerline{\includegraphics[width=1\textwidth]{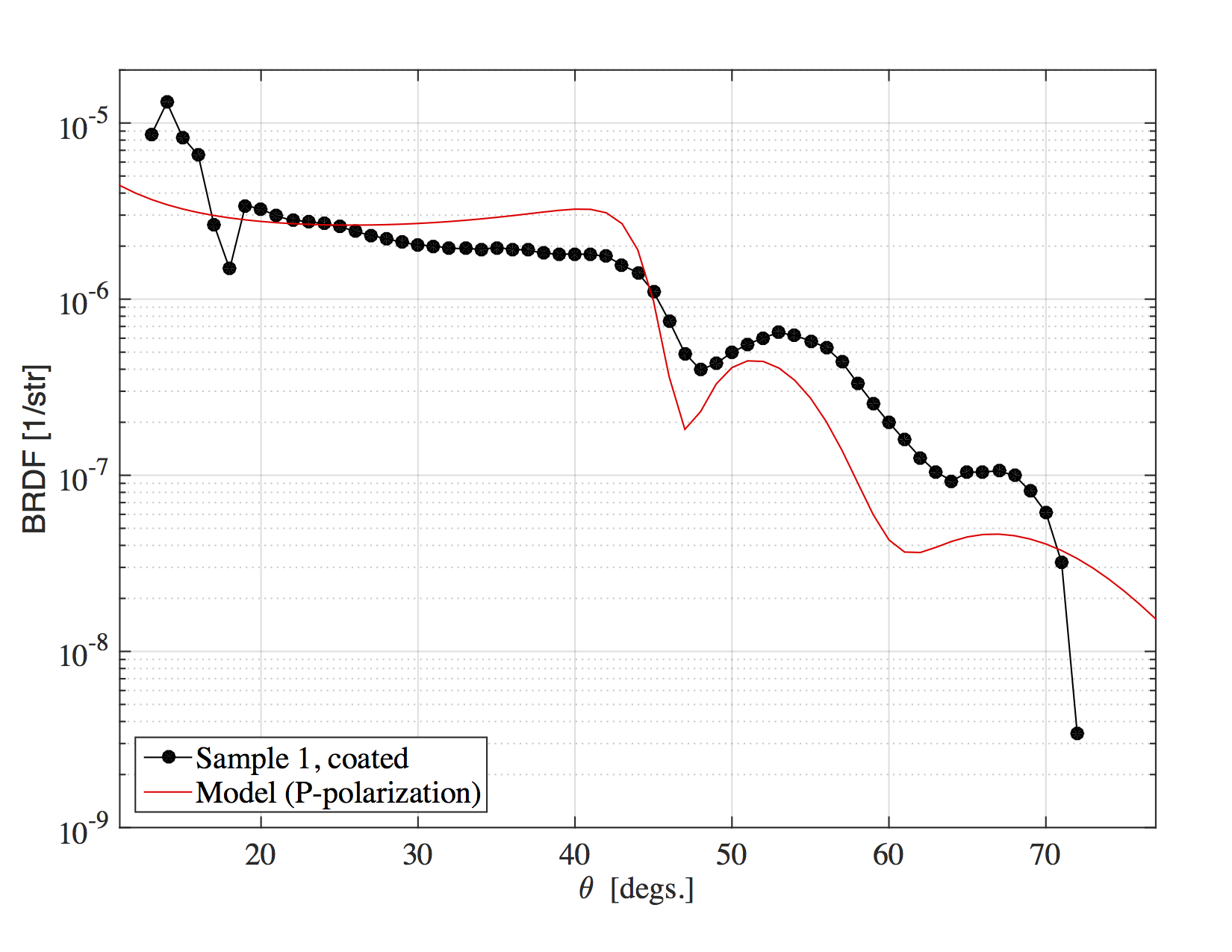}}
\caption{BRDF values versus scattering angle for Sample~1 after coating, compared to a model based on correlated roughness scattering at the coating interfaces.}\label{fig:measured-modeled}
\end{figure}

In order to better understand which features of the coatings led to the increase in BRDF and its angular features, a model was developed following references~\cite{Amra:93,Amra:94i,Amra:94ii}. The model coating was produced for Sample~1 following the designed coating layer thicknesses and materials obtained from the manufacturer. The modeled incident beam was 1064\,nm wavelength horizontally-polarized (p-) light at normal incidence and the scattered light detection was made for horizontal polarization, since first-order scattering (valid for polished surfaces) does not create cross-polarized scattering in the plane of incidence. Figure~\ref{fig:measured-modeled} shows the simulated scattered light for Sample~1 compared with the measured BRDF from Figure~\ref{fig:BRDFplot}. The angular position of the peaks is determined by the coating design, while the roughness spectrum gives the envelope of the BRDF and the optical properties of the simulated coating match closely the measured values. 

Only surface roughness of the layers was considered (i.e., no coating layer bulk scattering), and a perfect correlation of the interface roughness was assumed. Ideally, the measured BRDF of the uncoated sample would be fit to determine the roughness spectrum of the curved surface, and this would be used to predict the scattering of the dielectric coating, assuming that the coating layers follow the surface roughness exactly. However, the BRDF of the uncoated sample included scatter from the flat (back) surface, which was not superpolished, as described above, so fitting this data would overestimate the roughness power spectrum for the coating and thus the coated BRDF. For these reasons, the roughness spectrum was modeled as the sum of an exponential and a Gaussian function (as described in~\cite{Amra:93} with $\delta_e$ = 0.1\,nm, L$_e$ = 2000\,nm, $\delta_g$ = 0.5\,nm, L$_g$ = 150\,nm). 

Figure~\ref{fig:measured-modeled} indicates that interface scattering of the dielectric coating layers, given an exponential power spectrum, perfect correlation of interface roughness, and the design properties of the coating, could explain roughly the level and qualitative behavior of the coated BRDF measurements. No attempts were made to achieve better agreement between the measurement and the model because too many unknown parameters could be responsible for the discrepancies. These include thickness errors in the coating design, scattering from local defects, bias in the measurements, curvature of the substrate, and the quality of the polish. 

\section{Discussion and conclusion}

We estimate scattering losses of $5.00\pm.30$\,ppm and $3.38\pm.20$\,ppm for the highly-reflective coatings on Samples~1 and 2, respectively, over azimuthal angles of roughly $12^{\circ}-76^{\circ}$, corresponding to a solid angle around $1.5\pi$\,str. For state-of-the art superpolished optics, the wide angle range measured most often contributes the majority of the scatter losses (caused by small-scale roughness, point defects, particulates, and impurities). In a separate measurement of the optical loss of two very similar optics forming a long time storage cavity, the total round-trip loss was estimated to be 6-18\,ppm~\cite{Isogai:13}, comparable to these results. While this level of optical loss is considered low enough for the filter cavity currently being designed for Advanced LIGO~\cite{Evans:13}, reducing scattering losses could significantly improve LIGO's performance~\cite{Kwee:2014}. 

The model presented here indicates that the maxima and minima observed in the BRDF of the coated samples might arise from correlated interface scattering of the coating layers. This further suggests that lower coating scatter losses might be achieved with improved substrate polishing. Finally, it is worth noting that these scatter losses are somewhat higher than previously reported losses from highly reflective mirrors~\cite{Rempe:92, Magana-Sandoval:12} and from anti-reflection coatings~\cite{Padilla:14} (which are the lowest scatter measured with this setup) produced by other manufacturers.


\ack This work is supported by National Science Foundation Awards CAREER PHY-1255650 and RUI PHY-0970147 and by the Research Corporation for Science Advancement Cottrell College Science Award \#19839. We thank our colleagues in the LIGO Scientific Collaboration for fruitful discussions about this research and for review of this manuscript.

\section*{References}

\bibliographystyle{unsrt}
\bibliography{references_fc}

\begin{thebibliography}{10}

\bibitem{2013Sci...341.1215H}
N.~{Hinkley}, J.~A. {Sherman}, N.~B. {Phillips}, M.~{Schioppo}, N.~D. {Lemke},
  K.~{Beloy}, M.~{Pizzocaro}, C.~W. {Oates}, and A.~D. {Ludlow}.
\newblock {An Atomic Clock with 10$^{–18}$ Instability}.
\newblock {\em Science}, 341:1215--1218, September 2013.

\bibitem{Steinmetz2009}
T.~Steinmetz, T.~Wilken, C.~Araujo-Hauck, R.~Holzwarth, T.W. Hänsch, and
  T.~Udem.
\newblock Fabry-perot filter cavities for wide-spaced frequency combs with
  large spectral bandwidth.
\newblock {\em Applied Physics B}, 96(2-3):251--256, 2009.

\bibitem{2012PhRvD..85f4007H}
C.~J. {Hogan}.
\newblock {Interferometers as probes of Planckian quantum geometry}.
\newblock {\em Phys.Rev.D}, 85(6):064007, March 2012.

\bibitem{Willke:98}
B.~Willke, N.~Uehara, E.~K. Gustafson, R.~L. Byer, P.~J. King, S.~U. Seel, and
  Jr. R.~L.~Savage.
\newblock Spatial and temporal filtering of a 10-w nd:yag laser with a
  fabry--perot ring-cavity premode cleaner.
\newblock {\em Opt. Lett.}, 23(21):1704--1706, Nov 1998.

\bibitem{Araya:97}
Akito Araya, Norikatsu Mio, Kimio Tsubono, Koya Suehiro, Souichi Telada,
  Masatake Ohashi, and Masa-Katsu Fujimoto.
\newblock Optical mode cleaner with suspended mirrors.
\newblock {\em Appl. Opt.}, 36(7):1446--1453, Mar 1997.

\bibitem{Gossler:02}
S~Gossler, M~M Casey, A~Freise, H~Grote, H~Lück, P~McNamara, M~V Plissi, D~I
  Robertson, N~A Robertson, K~Skeldon, K~A Strain, C~I Torrie, H~Ward,
  B~Willke, J~Hough, and K~Danzmann.
\newblock The modecleaner system and suspension aspects of geo 600.
\newblock {\em Classical and Quantum Gravity}, 19(7):1835, 2002.

\bibitem{Fricke:12}
Tobin~T Fricke, Nicolás~D Smith-Lefebvre, Richard Abbott, Rana Adhikari,
  Katherine~L Dooley, Matthew Evans, Peter Fritschel, Valery~V Frolov, Keita
  Kawabe, Jeffrey~S Kissel, Bram J~J Slagmolen, and Sam~J Waldman.
\newblock Dc readout experiment in enhanced ligo.
\newblock {\em Classical and Quantum Gravity}, 29(6):065005, 2012.

\bibitem{Caves:85}
Carlton~M. Caves and Bonny~L. Schumaker.
\newblock New formalism for two-photon quantum optics. i. quadrature phases and
  squeezed states.
\newblock {\em Phys. Rev. A}, 31:3068--3092, May 1985.

\bibitem{buonanno2000optical}
Alessandra Buonanno and Yanbei Chen.
\newblock Optical noise correlations and beating the standard quantum limit in
  ligo-ii.
\newblock {\em Class. Quantum Gravity}, 18(gr-qc/0010011):L95--L101, 2000.

\bibitem{ligo:2011}
LIGO~Scientific Collaboration et~al.
\newblock A gravitational wave observatory operating beyond the quantum
  shot-noise limit.
\newblock {\em Nature Physics}, 7(12):962--965, 2011.

\bibitem{aasi:2013}
J~Aasi, J~Abadie, BP~Abbott, R~Abbott, TD~Abbott, MR~Abernathy, C~Adams,
  T~Adams, P~Addesso, RX~Adhikari, et~al.
\newblock Enhanced sensitivity of the ligo gravitational wave detector by using
  squeezed states of light.
\newblock {\em Nature Photonics}, 7(8):613--619, 2013.

\bibitem{PhysRevA.71.013806}
Simon Chelkowski, Henning Vahlbruch, Boris Hage, Alexander Franzen, Nico
  Lastzka, Karsten Danzmann, and Roman Schnabel.
\newblock Experimental characterization of frequency-dependent squeezed light.
\newblock {\em Phys. Rev. A}, 71:013806, Jan 2005.

\bibitem{Evans:13}
M.~Evans, L.~Barsotti, P.~Kwee, J.~Harms, and H.~Miao.
\newblock Realistic filter cavities for advanced gravitational wave detectors.
\newblock {\em Phys. Rev. D}, 88:022002, Jul 2013.

\bibitem{Kwee:2014}
P.~Kwee, J.~Miller, T.~Isogai, L.~Barsotti, and M.~Evans.
\newblock Decoherence and degradation of squeezed states in quantum filter
  cavities.
\newblock {\em Phys. Rev. D}, 90:062006, Sep 2014.

\bibitem{Fritschel:2014}
The LIGO~Scientific Collaboration.
\newblock Advanced ligo.
\newblock {\em Classical and Quantum Gravity}, 32(7):074001, 2015.

\bibitem{Harry:2011book}
Gregory Harry, Timothy~P. Bodiya, and Riccardo DeSalvo, editors.
\newblock {\em Optical Coatings and Thermal Noise in Precision Measurement}.
\newblock Cambridge University Press, 2011.

\bibitem{macleod2001thin}
Hugh~Angus Macleod.
\newblock {\em Thin-film optical filters}.
\newblock CRC Press, 2001.

\bibitem{0264-9381-21-5-083}
The~VIRGO Collaboration, F~Beauville, D~Buskulic, R~Flaminio, F~Marion,
  A~Masserot, L~Massonnet, B~Mours, F~Moreau, J~Ramonet, E~Tournefier,
  D~Verkindt, O~Veziant, M~Yvert, R~Barillé, V~Dattilo, D~Enard, F~Frasconi,
  A~Gennai, P~La Penna, M~Loupias, F~Paoletti, L~Bracci, G~Calamai, E~Campagna,
  G~Conforto, E~Cuoco, I~Fiori, G~Guidi, G~Losurdo, F~Martelli, M~Mazzoni,
  B~Perniola, R~Stanga, F~Vetrano, A~Viceré, D~Babusci, G~Giordano, J-M
  Mackowski, N~Morgado, L~Pinard, A~Remillieux, F~Acernese, F~Barone,
  E~Calloni, R~De Rosa, L~Di Fiore, A~Eleuteri, L~Milano, K~Qipiani,
  I~Ricciardi, G~Russo, S~Solimeno, M~Varvella, F~Bondu, A~Brillet,
  E~Chassande-Mottin, F~Cleva, T~Cokelaer, J-P Coulon, B~Dujardin, J-D
  Fournier, H~Heitmann, C~N Man, F~Mornet, J~Pacheco, A~Pai, H~Trinquet, J-Y
  Vinet, N~Arnaud, M~Barsuglia, M~A Bizouard, V~Brisson, F~Cavalier, M~Davier,
  P~Hello, P~Heusse, S~Kreckelberg, C~Boccara, V~Loriette, J~Moreau, V~Reita,
  P~Amico, L~Bosi, L~Gammaitoni, M~Punturo, F~Travasso, H~Vocca, L~Barsotti,
  S~Braccini, C~Bradaschia, G~Cella, C~Corda, A~Di Virgilio, I~Ferrante,
  F~Fidecaro, A~Giazotto, E~Majorana, L~Holloway, R~Passaquieti, D~Passuello,
  R~Poggiani, A~Toncelli, M~Tonelli, L~Brocco, S~Frasca, C~Palomba, P~Puppo,
  P~Rapagnani, and F~Ricci.
\newblock The virgo large mirrors: a challenge for low loss coatings.
\newblock {\em Classical and Quantum Gravity}, 21(5):S935, 2004.

\bibitem{Isogai:13}
T.~Isogai, J.~Miller, P.~Kwee, L.~Barsotti, and M.~Evans.
\newblock Loss in long-storage-time optical cavities.
\newblock {\em Opt. Express}, 21(24):30114--30125, Dec 2013.

\bibitem{Magana-Sandoval:12}
Fabian Magana-Sandoval, Rana~X. Adhikari, Valera Frolov, Jan Harms, Jacqueline
  Lee, Shannon Sankar, Peter~R. Saulson, and Joshua~R. Smith.
\newblock Large-angle scattered light measurements for quantum-noise filter
  cavity design studies.
\newblock {\em J. Opt. Soc. Am. A}, 29(8):1722--1727, Aug 2012.

\bibitem{Padilla:14}
Cinthia Padilla, Peter Fritschel, Fabian~Maga\ {n}a Sandoval, Erik Muniz,
  Joshua~R. Smith, and Liyuan Zhang.
\newblock Low scatter and ultra-low reflectivity measured in a fused silica
  window.
\newblock {\em Appl. Opt.}, 53(7):1315--1321, Mar 2014.

\bibitem{Stover:2012book}
J.~C. Stover.
\newblock {\em Optical Scattering}.
\newblock SPIE Press, 3rd edition, 2012.

\bibitem{Amra:93}
Claude Amra, C.~Gr\`{e}zes-Besset, and L.~Bruel.
\newblock Comparison of surface and bulk scattering in optical multilayers.
\newblock {\em Appl. Opt.}, 32(28):5492--5503, Oct 1993.

\bibitem{Amra:94i}
C.~Amra.
\newblock Light scattering from multilayer optics. i. tools of investigation.
\newblock {\em J. Opt. Soc. Am. A}, 11(1):197--210, Jan 1994.

\bibitem{Amra:94ii}
C.~Amra.
\newblock Light scattering from multilayer optics. ii. application to
  experiment.
\newblock {\em J. Opt. Soc. Am. A}, 11(1):211--226, Jan 1994.

\bibitem{Rempe:92}
G.~Rempe, R.~Lalezari, R.~J. Thompson, and H.~J. Kimble.
\newblock Measurement of ultralow losses in an optical interferometer.
\newblock {\em Opt. Lett.}, 17(5):363--365, Mar 1992.

\end{thebibliography}

\end{document}